\input harvmac
\input epsf

\newcount\figno
\figno=0
\def\fig#1#2#3{
\par\begingroup\parindent=0pt\leftskip=1cm\rightskip=1cm\parindent=0pt
\baselineskip=12pt
\global\advance\figno by 1
\midinsert
\epsfxsize=#3
\centerline{\epsfbox{#2}}
\vskip 14pt

{\bf Fig. \the\figno:} #1\par
\endinsert\endgroup\par
}
\def\figlabel#1{\xdef#1{\the\figno}}
\def\encadremath#1{\vbox{\hrule\hbox{\vrule\kern8pt\vbox{\kern8pt
\hbox{$\displaystyle #1$}\kern8pt}
\kern8pt\vrule}\hrule}}

\overfullrule=0pt

\noblackbox
\parskip=1.5mm

\def\Title#1#2{\rightline{#1}\ifx\answ\bigans\nopagenumbers\pageno0
\else\pageno1\vskip.5in\fi \centerline{\titlefont #2}\vskip .3in}

\font\caps=cmcsc10

\noblackbox
\parskip=1.5mm

  
\def\npb#1#2#3{{ Nucl. Phys.} {\bf B#1} (#2) #3 }
\def\plb#1#2#3{{ Phys. Lett.} {\bf B#1} (#2) #3 }
\def\prd#1#2#3{{ Phys. Rev. } {\bf D#1} (#2) #3 }
\def\prl#1#2#3{{ Phys. Rev. Lett.} {\bf #1} (#2) #3 }

\def\pr#1#2#3{{ Phys. Rev. } {\bf #1} (#2) #3 }

\def\bb#1{{[arXiv:hep-th/#1]}}

\def\heph#1{{[arXiv:hep-ph/#1]}}

\def\jhep#1#2#3{{ J. High Energy Phys.} {\bf #1} (#2) #3 }


\def\CM{{\cal M}}   
\def\CN{{\cal N}}


\def\dj{\hbox{d\kern-0.347em \vrule width 0.3em height 1.252ex depth
-1.21ex \kern 0.051em}}

\def\half{{1\over 2}\,}

\def\Tr{{\rm Tr\,}}
\def\tr{{\rm tr\,}}

\def\pt{\partial}

\def\Dirac{\,\raise.15ex\hbox{/}\mkern-13.5mu D}
\def\dirac{\,\raise.15ex\hbox{/}\kern-.57em \partial}
\def\aslash{\,\raise.15ex\hbox{/}\mkern-13.5mu A}

\def\shalf{{\ifinner {\textstyle {1 \over 2}}\else {1 \over 2} \fi}} 
\def\sshalf{{\ifinner {\scriptstyle {1 \over 2}}\else {1 \over 2} \fi}} 
\def\sfourth{{\ifinner {\textstyle {1 \over 4}}\else {1 \over 4} \fi}}

\lref\rlus{M. L\"uscher, \npb{219}{1983}{233.}P. van Baal, \bb{0008206.}}

\lref\rtoni{A. Gonz\'alez-Arroyo, \bb{9807108.}}

\lref\renri{E. Alvarez and A. Nieto, \prd{41}{1990}{3850.}} 

\lref\rtoy{N. Arkani-Hamed, S. Dimopoulos and S. Kachru, \bb{0501082.}}

\lref\rhoso{Y. Hosotani, \plb{126}{1983}{309.}}

\lref\rhosn{Y. Hosotani, \heph{0504272.}}

\lref\rgpy{D. Gross, R. Pisarski and L. Yaffe, Rev. Mod. Phys. {\bf 53} (1981)
43.}

\lref\rbanks{T. Banks, \bb{0412129.}}

\lref\rlan{R. Bousso and J. Polchinski, \jhep{0006}{2000}{006}  
\bb{0004134.} M. Douglas, \jhep{0305}{2003}{046}  
\bb{0303194.}} 

\lref\rdec{N. Arkani-Hamed, A.G. Cohen and H. Georgi, \prl{86}{2001}{4757}
\bb{0104005.}}

\lref\rtof{G. 't Hooft, \npb{153}{1979}{141.}}

\lref\ranton{E. Alvarez and A. F.  Faedo, \bb{0602150.}}  

\lref\raha{O. Aharony,  
 J. Marsano, S. Minwalla, K. Papadodimas, M. Van Raamsdonk and  T. Wiseman,  
\jhep{0601}{2006}{140}  
\bb{0508077.}} 

\lref\rkklt{S. Kachru, R. Kallosh, A. Linde and S. Trivedi, \prd{68}{2003}
{046005} 
\bb{0301240.}}

\lref\rsus{L. Susskind, \bb{0302219.}}

\lref\rus{ 
J.L.F. Barb\'on and C. Hoyos, \bb{0507267.} C. Hoyos, \bb{0512303.} 
}

\lref\rori{A. Armoni, M. Shifman and G. Veneziano, \npb{667}{2003}{170} 
\bb{0302163.}} 

\lref\rwittt{E. Witten, \bb{9712028.}} 

\lref\rind{E. Witten, Adv. Theor. Math. Phys. {\bf 5} (2002) 841, 
\bb{0006010.}} 

\lref\rmat{J. Gomis, F. Marchesano and D. Mateos, \jhep{0511}{2005}{021}  
\bb{0506179.}} 

\lref\rtoyy{C. Escoda, M. G\'omez-Reino and F. Quevedo, 
JHEP {\bf 0311} (2003) 065, 
\bb{0307160.}  K. R. Dienes, E. Dudas and T. Gherghetta, 
\prd{72}{2005}{026005}  
\bb{0412185.}}


\baselineskip=15pt

\line{\hfill IFT UAM/CSIC-2006-09}
\line{\hfill {\tt hep-th/0602285}}

\vskip 1.0cm

\Title{\vbox{\baselineskip 12pt\hbox{}
 }}
{\vbox {\centerline{Dynamical Higgs potentials with a landscape    }
\vskip10pt
\centerline{   }
}}

\vskip1.0cm

\centerline{$\quad$ {\caps 
J.L.F. Barb\'on
 and   
C. Hoyos 
}}
\vskip0.7cm

\centerline{{\sl  Instituto de F\'{\i}sica Te\'orica IFT UAM/CSIC }}
\centerline{{\sl  C-XVI,  
 UAM, Cantoblanco 28049. Madrid, Spain }}
\centerline{{\tt jose.barbon@uam.es, c.hoyos@uam.es}}

\vskip0.2cm


\vskip1.0cm

\centerline{\bf ABSTRACT}

 \vskip 0.3cm

 \noindent 

We consider one-loop effective potentials for adjoint Higgs fields
that originate from   flat holonomies in  toroidal compactification of
gauge theories.  
We show that such potentials are ``landscape-like" for large gauge groups
and generic non-supersymmetric  
matter representations. 
In particular,  there is a large number of vacua with similar local
properties,   scanning   a broad
band of vacuum energies. 

\vskip 1.5cm

\Date{February 2006}

\vfill





\baselineskip=15pt

\newsec{Introduction}

\noindent

In discussions of effective  potentials defined  over moduli spaces of 
string compactification 
 (following the blueprint of \refs\rkklt),
one  often  emphasizes  their  very intricate nature.  Working in the 
framework of low-energy effective actions (a notoriously dubious procedure
for the task at hand \refs\rbanks),    
the introduction
 of various   trapped fluxes and other  nonperturbative effects  
 generates  effective potentials 
with  very rich structure, depicting an almost infinite
variety of minima, maxima, valleys and rifts dubbed ``the landscape" 
 \refs\rsus. 
The complicated features of these effective potentials are largely a consequence
of the discreteness introduced by a large number of quantized fluxes and
the variety of possible brane and singularity configurations \refs\rlan.
 In a sense,
it comes as no surprise that a very complicated potential should be obtained
 from a 
system with a rich cocktail of degrees of freedom.

However, the scanning of low-energy couplings in a broad band of values, 
particularly the vacuum energy, is considered to be
 a general feature of landscape
potentials. For example, it is incorporated as a crucial ingredient in 
 field-theoretical toy examples \refs{\rtoyy, \rtoy}.   
In this brief note we study one such field-theoretical toy model in which a
 landscape-like 
potential  is generated {\it dynamically} from a higher dimensional gauge
theory.  

 In this letter, we will
 refrain from any discussion of possible phenomenological 
applications, except for some brief comments at the end. Our main motivation
is to show how a very complicated-looking potential can be generated out
of relatively simple  systems. In particular, we have few adjustable
parameters  affecting significantly the structure of the potentials, whose
properties are largely a consequence of group theory effects. 

We consider perturbative effective potentials for adjoint Higgs fields that
are generated by dimensional reduction on a torus with flat connections. 
This type of geometrical construction is very well known, and has been widely
studied since \refs\rhoso\ (see \refs\rhosn\ for a recent summary). The
 essential ingredient responsible for the ``landscape" features
is a parametrically large rank of the  gauge group
 and the breaking of a global discrete symmetry
that maps out vacua with similar local properties.

\newsec{Effective potentials  for  flat connections}

\noindent

We begin by reviewing some standard facts about adjoint Higgs fields generated
upon toroidal compactification. 
Let us consider
a  Yang--Mills model in a $(d+n)$-dimensional spacetime
of the form ${\bf R}^d \times {\bf T}^n$, a product of a standard flat
$d$-dimensional Minkowski spacetime and an internal torus of size $L$, which
we take orthogonal for simplicity. The
classical action
\eqn\yma{
S_h = {1\over 2g_h^2} \int_{{\bf R}^d \times {\bf T}^n} \tr\,|F_{AB}|^2 
}
reduces, for distance scales much larger than $L$, to the effective
system
\eqn\ymea{
S= 
{1\over 2g^2} \int_{{\bf R}^d} \tr\,\left(|F_{\mu\nu}|^2 + |D_\mu \Phi_a\,|^2 -
 \left|\,
[\Phi_a , \Phi_b\,]\,\right|^2
+\dots\right)
\;,}
where we have split the $(d+n)$-dimensional indices $(A,B, \dots)$ into 
${\bf R}^d$ spacetime indices $(\mu,\nu, \dots)$
 and internal ${\bf T}^n$  indices $(a,b,\dots)$.
Each field  $\Phi_a$ is an adjoint Higgs  that originates in the Fourier
modes of the gauge field components along the $n$-torus and the commutator
term is their classical potential, coming from the ``magnetic" energy
density, $F_{ab}F^{ab}$, 
of field-strengths  along the torus directions. The dots stand for 
higher order terms suppressed by powers of $EL$ in a low-energy expansion
with $E\ll 1/L$, and the classical mapping between high and low energy
 couplings is the
standard Kaluza--Klein relation $g_h^2 = L^n g^2$. Only the zero modes of the gauge field
on ${\bf T}^n$ survive as {\it massless} adjoint Higgses in the effective theory
below the gap $1/L$. 
These are the flat connections on ${\bf T}^n$, defined by the vanishing of
the magnetic energy $F_{ab}=0$. A  gauge-covariant description
of these degrees of freedom is given by topologically nontrivial Wilson lines
wrapped around the noncontractible cycles of ${\bf T}^n$. 

In general, the space of flat connections on ${\bf T}^n$ has
 connected components. These can be the result of  specific  
group-theory properties  (e.g. the    nontrivial
 commuting triples for $n=3$ and particular gauge groups  \refs\rwittt),
or       be associated to    
 topological structure of the  gauge bundles, such as
  non-abelian  electric and magnetic fluxes  
through the torus \refs\rtof. Each connected component of the moduli
space admits a parametrization in terms of constant commuting
 gauge connections in
some appropriate subgroup of the gauge group. Since most of our results apply to
each connected component separately,  
 in the following we focus on   the particular  case of the single connected
component containing the identity Wilson lines,   
 corresponding to bundles admiting
 periodic boundary conditions on the torus.  A   
convenient parametrization of these flat connections is given by 
gauge fields that are constant on  
${\bf T}^n$ and lie on the Cartan subalgebra 
of the gauge group. Modding out by the remaining gauge transformations one
finds the moduli space
\eqn\mod{
\CM = {({\bf T_{\rm C}}^r\,)^n \over W}
\;,}
where $r$ is the rank of the gauge group and $W$ is the Weyl group that
permutes the eigenvalues of the Wilson lines.
 We refer to the  torus
$({\bf T_{\rm C}}^{r})^n$ as the ``toron valley",  defined as
the product of $n$ copies of  ${\bf R}^{r}/2\pi {\widetilde \Lambda}_{\rm r} $, where
${\widetilde \Lambda}_{\rm r}$ is the dual of the root lattice.
A set of  coordinates on $\CM$ is provided by the constant  commuting
connections
\eqn\dph{
L {\vec \Phi} \equiv {\vec \phi} =
\sum_{i=1}^{r} {\vec \phi}_i \,H_i
\;,}
 with $H_i$ a complete set of generators
 of the Cartan subalgebra. In these coordinates, the Cartan torus 
 identification  on  
 ${\vec \phi}$  consists of   
translations by   $4\pi \alpha\, {\vec n}$, with $\alpha$ a root and 
${\vec n} \in {\bf Z}^n$ a
vector
of integers.
 The Weyl group acts by reflections on the hyperplanes orthogonal to the
 roots, and makes $\CM$ into an orbifold.

In the effective gauge theory
on ${\bf R}^d$, the moduli space $\CM$ is
to be regarded as the target space for the Higgs fields. For $d>2$, it
is also  the space
of their possible ``expectation values" in a Coulomb phase, i.e. the
gauge group left unbroken by the Higgs mechanism at a generic point on  
$\CM$ is the  maximal abelian subgroup. On 
submanifolds of higher codimension 
 we have enhanced gauge symmetry, which remains
completely unbroken at the origin of moduli space, together with 
the points on $\CM$ that are obtained from ${\vec \phi}=0$ by the 
action of twisted gauge transformations \refs\rtof.
 These are transformations $U({\vec z}\,)$ that act by 
finite shifts of $\phi$, and 
are twisted around the nontrivial cycles of the torus as
\eqn\lgt{
T_a \,U({\vec z}\,)\, T_a^{-1} = z_a \,U({\vec z}\,) 
\;,}
where $T_a$ is a translation by $L$ on the $a$-th direction of the torus
and $z_a$ is a phase in the center of the gauge group. For example, for
$SU(N)$ the center is ${\bf Z}_N$ and $z_a$ is given by any $N$-th root
of unity. In general, if $N$ is the cardinal of the center, we find
$N^n$ vacua that are locally identical to the unbroken vacuum labelled by
${\vec \phi} =0$. 

\subsec{Effective potentials}

\noindent

The low-energy 
effective action for fields, ${\vec \phi}$, 
 parametrizing the Cartan torus (i.e. for those
with vanishing classical potential $[\Phi^a, \Phi^b\;]=0$) includes an
effective potential induced by integrating out all the high-energy
degrees of freedom. Working in  perturbation theory in the  Yang--Mills 
coupling, we have a natural expansion  parameter at the compactification 
scale $1/L$. For a rank $N$ gauge group,  it is  given by the 
dimensionless 't Hooft coupling  
   $\lambda=g^2 (L) \,N \, L^{4-d} $, where $g^2 (L)$ is the effective
$d$-dimensional Yang--Mills coupling,  renormalized at
the scale $1/L$.  This parameter, defined at the matching scale $1/L$ runs
with energy  according to  
$$
\lambda_{\rm eff} (E) = \lambda \; (EL)^{d_{\rm eff}-4}
\;,$$
with logarithmic running for $d_{\rm eff}=4$. The effective dimension
is given by $d_{\rm eff} = d$ for $EL<1$,
 whereas $d_{\rm eff}=d+n$ for $EL>1$.  
 Of course, perturbation theory necessarily breaks down at  sufficiently high  
energies for $d+n >4$. The strong-coupling threshold is defined by $
\lambda_{\rm eff} (\Lambda_{\rm UV}) = 1$, or   
\eqn\thrsld{
\Lambda_{\rm UV}= {  \lambda^{-{1\over d+n-4}} \over L}  
\;,} 
and defines the scale beyond which 
 some ultraviolet (UV) completion must be provided. 
Effects of such UV physics decouple below the compactification scale 
$L$ as inverse powers of $\Lambda_{\rm UV}\, L$, or equivalently, 
as positive fractional powers of $\lambda$.

Keeping in mind these limitations of the perturbative approach, the
one-loop effective potential can be defined by the gaussian functional
integral over those gauge fields on  ${\bf R}^d \times {\bf T}^n$ that are
  orthogonal
to the flat connections. Since the flat connections ${\vec \phi}$ are
not integrated over, they  function as a background field and we may use
the standard machinery of the background field gauge.  
We obtain for the one-loop effective potential
\eqn\bfg{
\int_{{\bf R}^d} V_{\rm eff} ({\vec \phi}\,)
 = \half \Tr'_{{\cal V}^T} \; \log \; \left(\;-D_M D^M\;
\right)_{{\bf R}^d \times {\bf T}^n}  - \left[{\vec \phi}=0\right]\;,
}
where we have normalized the potential to vanish 
at the origin of moduli space ${\vec \phi} =0$ (here and in the following,
the  square bracket notation  means that we subtract the previous expression
with the appropriate vanishing parameter).  The
adjoint covariant derivative operator $D_M = \pt_M + i \left[A_M, \;\;\right]$
 is evaluated at the
flat connection $A_M = (0, {\vec \Phi}\,)$. The trace in \bfg\ is taken
over the space ${\cal V}^T$ of transverse vector fields on    
  ${\bf R}^d \times {\bf T}^n$, and the prime means that zero modes of
the operator $-D_M D^M$ are left out of the trace.

Examples of such potentials have been extensively studied in the literature.
The cases $d=4$ have been considered repeatedly as models for
Higgs sectors, starting with \refs\rhoso. The cases $n=1$ are well-known
from the literature on thermal field theory \refs\rgpy\ (see the recent
\refs\raha\ for the case $d=0$),
 while the case $d=1$ appears
  in the small-volume expansion of gauge theories on tori \refs{\rlus,
\rus}.\foot{In fact, the ``landscape-like" properties reported below were
originally found in the particular cases studied in \refs\rus.}

Factoring out the Lorentz and gauge group components of the functional trace we find
\eqn\efp{
 V_{\rm eff} ({\vec \phi}\,) = 
(d+n-2) \sum_{\alpha \in {\rm roots}}  V_0\left(\,\alpha \cdot {\vec \phi}\;
\right)
\;,}
where the dot stands for the Cartan inner product, $\alpha \cdot {\vec \phi}=
\sum_{i=1}^r \alpha_i {\vec \phi}_i$. We have used the fact that the flat connection 
lies on the Cartan subalgebra, and the roots are the eigenvalues of the Cartan
generators.    The  purely scalar and abelian function, $V_0$, is given by 
$$
\int_{{\bf R}^d} V_0 ({\vec \xi}\,) = \half \Tr\;\log\; \left[-\pt_\mu \pt^\mu
 -\left({\vec \pt} + i
\;{{\vec \xi}\over L}\;\right)^2 \,\right]_{{\bf R}^d \times {\bf T}^n}
-\half 
 \Tr\;\log \;\left[-\pt_\mu \pt^\mu - {\vec \pt}^{\;2} \right]_{{\bf R}^d 
\times {\bf T}^n}
$$
where the differential operators act now over  scalar functions on
${\bf R}^d \times {\bf T}^n$, and
 the determinants are calculated in Euclidean signature. 
A proper time representation of this determinant is the following
\eqn\ptime{
V_0 \left({\vec \xi}\;\right) = -\half
 \int_0^\infty {dt \over t} \int{d^d p \over
(2\pi)^d} \sum_{{\vec n}\in {\bf Z}^n}  e^{-t\,\left(p^2 +  \left({2\pi  
\vec n \over L} +
 {{\vec \xi} \over L} \,\right)^2 \right)} - \left[\; {\vec \xi} =0\; \right]
\;.}
We can obtain a physical interpretation of the potential by integrating over
the time component of momenta in  ${\bf R}^d$, 
\eqn\cw{
V_{\rm eff}\left({\vec \phi}\;\right)
 = \half (d+n-2) \sum_{\alpha} \sum_{{\vec n}\in {\bf Z}^n}
 \int{d{\bf p} \over
(2\pi)^{d-1}} \; \sqrt{{\bf p}^2 + \left({{\vec n}\over L}
 + {\alpha \cdot {\vec \phi} 
\over 2\pi L} \right)^2 }-
\left[ \,{\vec \phi} =0\;\right] 
\;,}
which has a straightforward interpretation as the sum of zero-point energies of
all field oscillators for each of the  $d+n-2$ spin degrees of freedom. 
Field oscillators along the compact torus suffer a frequency shift due to the
constant holonomy ${\vec \phi}$. Therefore, the effective potential is  of
Coleman--Weinberg type.   

As expected from its definition,  the subtracted
potential \cw\ gets no contribution from those  modes in the Cartan subalgebra
that are also constant throughout the 
torus,  
  corresponding  to the ${\vec n}=0, \alpha=0$ terms in the
sums. Since these modes are precisely given by the background flat connections,
 ${\vec \phi}$, we can view \cw\ as a Wilsonian potential in which all non-zero modes plus
non-abelian constant modes on the torus are integrated out. In fact, the
constant   non-abelian modes at 
 ${\vec n}=0$ 
 have an effective mass of order $|\alpha\cdot {\vec \phi}\,| /L$, and
 the Wilsonian separation of scales breaks down at points
in $\CM$ where the Higgs mechanism turns off, i.e. the orbifold points. The  
extra massless degrees of freedom induce infrared singularities in \cw. 
In order to estimate these,  
 we examine the
 ${\vec n}=0$ terms, with an  ultraviolet cutoff in place,
$$
\Delta V({\vec \xi}\,)_{\rm IR} = -{1\over (4\pi)^{d/2}} \int_{\epsilon^2}^{
\infty} {dt \over t} t^{-d/2} \,e^{-t
\,{\vec \xi}^{\;2} /L^2}
\;.$$
In this expression, it is plain that $|{\vec \xi}| /L$ plays the
role of an infrared cutoff in the proper time integral, which is itself
regularized  in
the ultraviolet by the  length-squared  $\epsilon^2$. We can
choose $\epsilon = L$ to consider the contribution of non-abelian constant
modes with energies ranging from the compactification scale $1/L$ down to
the infrared cutoff $|{\vec \xi}| /L$. Then we may write the previous expression
in the form
$$
\Delta V({\vec \xi}\,)_{\rm IR} = -{1\over (4\pi)^{d/2} L^d} \;
\left({\vec \xi}^{\;2}\right)^{d/2} \;\Gamma\left({\vec \xi}^{\;2}, -d/2\right)
\;,$$
in terms of the incomplete Gamma function. For odd values of $d$ we find
a non-analytic behaviour of branch-point type, 
  $V_0 ({\vec \xi}\,) \sim ({\vec \xi}^{\;2}\,)^{d/2}$, with
 a logarithmic correction $V_0 ({\vec \xi}\,)
 \sim ({\vec \xi}^{\;2}\,)^{d/2}
 \;\log\,({\vec \xi}^{\;2})$ for even values of $d$.
In particular, we have a conical singularity for $d=1$, the case studied
in \refs{\rlus, \rus}.
At any rate,  as long as  $d>2$ the non-analytic  terms are
 quantitatively subleading to
the analytic terms of order ${\vec \xi}^{\;2}$, 
 induced by the high-energy modes.

An alternative  representation of the effective potential
  is obtained by first integrating
over the full ${\bf R}^d$ momenta and subsequently performing a Poisson 
resummation on the discrete sum over the ${\bf Z}^n$ lattice,
\eqn\pr{
V_0\left({\vec \xi}\;\right)= -{L^n \over 2 (4\pi)^{d+n \over 2}} \int_0^\infty
{dt \over t} \;t^{-{d+n \over 2}} \sum_{{\vec \ell} \in {\bf Z}^n} 
e^{-{L^2 \over 4t} \, {\vec \ell}^{\;2}} \left( e^{i{\vec \ell} \;{\vec \xi}}
-1 \right)
\;.}
Notice that the subtraction of the ${\vec \xi}=0$ term can be incorporated as
a restriction in the discrete sum to non-vanishing 
${\bf Z}^n$ 
 vectors ${\vec \ell}\neq 0$. The discrete vectors ${\vec \ell}$ can be 
interpreted as ``winding" numbers around the torus ${\bf T}^n$ of 
first-quantized paths of the particles being integrated out. 
 Then, carrying out the proper-time integral
we arrive at
\eqn\finxi{
V_0\left({\vec \xi}\;\right) = {\Gamma\left({d+n \over 2}\right)
 \over \pi^{d+n \over 2}}  
 \;{1\over L^d} \;\sum_{{\vec \ell} \neq 0} {\sin^2 \left(
\shalf {\vec \ell}\; {\vec \xi}\;\right) \over \left({\vec \ell}^{\;2} 
\right)^{d+n \over 2}}
\;.}
  In this form the effective potential is seen to be positive definite, 
vanishing at the origin, with periodicity ${\vec \xi} \rightarrow {\vec \xi} +
2\pi {\bf Z}^n$, and reflection symmetry ${\vec \xi} \rightarrow -{\vec \xi}$.
The maxima of $V_0({\vec \xi}\;)$ sit at $(\pi, \pi, \dots, \pi)$ modulo 
$2\pi {\bf Z}^n$, whereas the minima are the images of ${\vec \xi} =0$ 
 under the periodicity 
symmetry.

An interesting property of the one-loop effective potential is its 
ultraviolet finiteness,
after a ${\vec \phi}$-independent constant  is appropriately subtracted. 
This is actually true to all orders in perturbation theory. The reason is
that any counterterm of the $(d+n)$-dimensional gauge theory must be 
gauge-invariant and, as such, a polynomial in covariant derivatives and
curvature field strengths. All those counterterms vanish on the space of
flat connections, and therefore $V_{\rm eff}$ cannot get UV divergences at
any order in perturbation theory (see \refs\ranton\ for a recent analysis
of counterterms in compactified gauge theories).
 The error made by extending the momentum
integrals and sums beyond the UV threshold $\Lambda_{\rm UV}$ vanishes as 
$\Lambda_{\rm UV}\, L$ goes to infinity. Therefore, all corrections to $V_{\rm eff}$
from the physics of the UV completion are controlled by the small parameter
$\lambda \ll 1$. 

\subsec{Generalizations}

\noindent

The function $V_0({\vec \xi}\;)$ also 
determines the effective potential induced by integrating out matter fields
in an arbitrary representation of the gauge group. For 
 scalar fields or fermions in an irreducible representation $R$, we must
substitute  the number of on-shell degrees of freedom, $n+d-2$ in 
\cw, by the corresponding number of on-shell 
degrees of freedom (polarization, charge, etc) in bosons or fermions,
${\cal N}_R$. In general, this number will also depend on whether the
fields in question are massive or massless. 
At the same time, we must substitute the sum over roots (i.e. the weights of
the adjoint representation) by a sum over weights in the representation $R$.
 At the end, we find 
\eqn\genp{
V_{\rm eff} \left({\vec \phi}\;\right) = \sum_R (-1)^{F_R} \;{\cal N}_R
 \; \sum_{\mu 
\in R} V_{M_R} \left(\mu \cdot {\vec \phi}\;\right) 
\;,}
where $(-1)^{F_R}$ is the fermion number of the representation. The
potential function $V_{M}$ is the massive generalization of \ptime, 
obtained by the simple replacement $p^2 \rightarrow p^2  + M^2$. 
Carrying out the substitution, the representation \pr\ is simply modified
by the insertion of $e^{-M^2 t}$ in the proper-time integral. Finally, the
final form of the potential \finxi\ gets modulated as
\eqn\finm{
V_M \left({\vec \xi}\;\right) ={2\over L^d} \;
\left({ML \over 2\pi}\right)^{d+n \over 2} \; \sum_{{\vec \ell} \neq 0} {
K_{d+n \over 2} \left(ML \sqrt{{\vec \ell}^{\;2}}\;\right)
 \over \left({\vec \ell}^{\;2}
\right)^{d+n \over 4}} \; \sin^2 \left(\shalf {\vec \ell} \; {\vec \xi}
\;\right)\;,
}
where $K_\nu (x)$ is a modified Bessel function of the second kind. In the
massless limit $ML \rightarrow 0$, we recover \finxi\ through the 
limit $K_\nu (x\rightarrow 0) \sim \shalf \Gamma(\nu) (2/x)^\nu$.  On the
other hand, the asymptotics
$K_\nu (x\gg 1) \sim e^{-x} \sqrt{\pi/2x}$ implies that the sum over  
 winding modes is effectively cutoff at $|{\vec \ell}\,| < (ML)^{-1}$. In
particular, for large masses on the compactification scale, $ML \gg 1$,
the whole potential is suppressed by a factor of $\exp(-ML)$.  

An interesting particular case concerns models with softly broken 
supersymmetry, i.e. broken by the mass splittings $M_R$.  In this case,
the total potential gets no contribution from the far UV regime,   
and the boson-fermion cancellation is complete when the non-supersymmetric
mass splittings are removed. In any case, notice that the mass terms do
not have a large effect on 
 the qualitative form of the potential function 
\finm, which is dominated by small values of $|{\vec \ell}|$ even for
the $M=0$ case. Therefore, the main effect of soft-breaking masses is a global
quenching of the associated effective potentials, without major
modifications of the qualitative features such as the symmetry properties (i.e. 
 the  location of the
vacua with unbroken gauge symmetry).  

 A point of detail concerns
the proper Wilsonian interpretation of the effective potentials induced by 
arbitrary matter representations. In the adjoint representation, all flat
connections are zero modes of the adjoint covariant derivative operator. 
 These zero modes were eliminated  from the partition sums
by  the vacuum-energy subtraction. For generic matter representations, zero
modes are located on submanifolds 
of zero measure in the moduli space ${\cal M}$.
Therefore, we integrate out all matter degrees of freedom and let singularities
of $V_{\rm eff}$ appear on ${\cal M}$, much in the same fashion as the
enhanced-symmetry 
 singularities  appear for the case of the adjoint representation.

Some further generalizations concern different spin structures for
fermions on ${\bf T}^n$ (which amount to  additive shifts of  
${\vec \phi}$ in all formulae, see for example \refs\renri),
 and twisted boundary conditions  for the
gauge bundles on ${\bf T}^n$. In this last case, the moduli space of flat
connections  is
partially  lifted (see \refs\rtoni\ for a review).
 On the remaining moduli space the analysis
is equivalent to what is described here.

\newsec{The Cartan--Weyl landscape}

\noindent

The potential 
\genp\ has ``landscape-like" features for generic representations.
The overall features are dictated by the crystallographic nature of the
compact moduli space $\CM$, in particular  the 
slightly different action of the Weyl group, $W$, 
 depending on  the representation under consideration.  
Associated to a given  weight $\mu$ in a general representation,
 there are codimension
$n$ hyperplanes on $\CM$, defined by   
$$
\mu \cdot {\vec \phi} = 0 \;{\rm mod}\; 2\pi{\bf Z}^n\;
,$$
that are fixed by $W$. 
Such Weyl hyperplanes host local minima of bosonic potentials (along
transverse directions) and local maxima of fermionic potentials. Therefore,
local minima (maxima) are found at intersections of Weyl hyperplanes, i.e.
at the edges of the so-called Weyl chamber. 
The effective potential contributed by a representation $R$ has structure
down to the overall scale determined by the  size of the Weyl chamber, 
inversely proportional to the norm of the highest weight.  
 Hence, very large representations induce potentials with 
short-distance scale  on $\CM$. The overall picture is that of an intricate
pattern of rifts and valleys along the Weyl hyperplanes associated to 
different representations.  

Fermionic contributions to the total effective potential look like ``inverted"
bosonic potentials. Hence, the singular locus of fermionic contributions is 
associated with local maxima rather than minima. Conversely, local minima
of fermionic potentials are related to local maxima of bosonic potentials,
and they could be smooth, just like the local maxima of the function
\finxi. Such smooth local minima of the full potential will typically 
break the gauge symmetry in a complicated pattern.

An interesting question is the fate of  vacua with ``unbroken" gauge symmetry, 
corresponding to the zeros of the pure Yang--Mills effective potential. If the
matter potential happens to be sufficiently 
smooth at those minima, its main effect is simply
to lift the unbroken vacuum to a non-vanishing vacuum energy.   
The surprising fact is  that these lifted vacua typically scan in a broad band
of energies.

We now focus on $SU(N)$ gauge theories, which give rise to  a very
large group of central conjugations, i.e.  $({\bf Z}_N)^n$ with cardinal
$N^n$.  In this case, we have a large number of unbroken symmetry vacua
on $\CM$ and we can enquiry how are they lifted by the matter potentials.  
The zeros of the potential \cw\ are solutions of   the equation
\eqn\eqzero{
\alpha \cdot {\vec \phi}_0 =0 \;\;{\rm mod}\;\;2\pi{\bf Z}^n
}
for any root $\alpha$. The $SU(N)$ roots can be written as $\alpha^{(ij)} =
\nu^i - \nu^j$, with $\nu^i$ the weights of the defining fundamental 
representation of $SU(N)$. 
These weights  satisfy $\sum_{i=1}^N \nu^i =0$ and 
$\nu^i \cdot \nu^j = (N\delta^{ij}-1)/2N$. Hence, any $N-1$ of them  
form a basis of the Cartan subalgebra. This basis {\it not} being 
exactly orthogonal will
prove a crucial fact in what  follows. 
Expressing \eqzero\ in this basis, we obtain the full set of solutions 
parametrized as  the discrete lattice
\eqn\dla{
{\vec \phi}_0 = 4\pi \sum_i {\vec n}_i \;\nu^i\;, \;\;\;{\vec n}_i \in {
\bf Z}^n
\;.}

In the pure Yang--Mills theory, all these vacua with unbroken gauge symmetry
are strict copies of the zero-Higgs vacuum ${\vec \phi}=0$. In fact, their
occurrence is related to the action of the $({\bf Z}_N)^n$ group of 
central conjugations, associated to large gauge transformations as in
\lgt. Perturbatively, these vacua remain disconnected, since the tunneling
amplitude across the potential barrier in $V_{\rm eff}$ is nonperturbative in
the gauge coupling. 

In the presence of additional matter degrees of freedom, these vacua get
lifted according to the value of 
$$
\sum_R (-1)^{F_R} \;\CN_R \;\sum_{\mu\in R} V_{M_R} \left(\,\mu\cdot {\vec \phi_0}\,
\right)
\;.$$
Thus, for a given representation, we must evaluate 
$$
V_M\left(4\pi \sum_i {\vec n}_i \;\mu\cdot \nu^i \;\right)
$$
for all weights $\mu$.  In general, any weight can be found in the weight lattice generated by the $\nu^i$. Therefore, the argument of the scalar function 
$V_M(
{\vec \xi}\,)$ will be determined by integer linear combinations of the scalar
products 
$$
4\pi \,\nu^i \cdot \nu^j = 2\pi \,\delta^{ij} - {2\pi \over N}
\;.$$
The term proportional to the Kronecker delta has no effect by the periodicity
properties of  \finxi, and we are left with a sum of terms of the form
$V_M(2\pi {\vec K}/N)$, where ${\vec K}$ is a ${\bf Z}^n$--valued vector defined
modulo $N$. If ${\vec K}$ is of $O(1)$ in the large $N$ limit, then the full
potential at this particular local vacuum is a sum of terms of $O(1/N^2)$, since
the potential can be considered  quadratic near the origin  
(for $d>2$). On
the other hand, if  ${\vec K} = O(N)$, we have a sum of terms of $O(1)$. 
The final scaling of the potential depends in each case on the multiplicity
from the sum over weights.

Our considerations refer only to the  energy shift of the $N^n$
 ``unbroken"
vacua in a $({\bf Z}_N)^n$ representation. 
In principle, the matter potential can alter the local properties of the
vacua beyond a simple shift of vacuum energy. If the slope of $V_R$ is
large enough at   
${\vec \phi_0}$, the local minimum can  disappear. 
 This is more likely to happen the larger is the representation
contributing to $V_{\rm eff}$, because the overall scale of the potential
is proportional to the dimension of the representation. For this reason,
stability of the unbroken vacua will require in general that the mass
of matter representations be sufficiently large, so that the matter
contributions are appropriately quenched by the $\exp(-ML)$ suppression
factor. For $ML\gg 1$ and near the origin, we may approximate
\eqn\apot{
V_M \left({\vec \xi}\;\right) \approx \Delta   \;
{\vec \xi}^{\;2} 
\;,}
with 
$$
\Delta \sim {(ML)^{d+n-1 \over 2} \over L^d} \;\exp(-ML)
\;.$$
For $d=2$ there is a correction by a logarithmic factor, $\log |{\vec \xi}\;|$,
whereas for $d=1$ the leading approximation is linear in $|{\vec \xi}\;|$.
In the following we consider the case $d>2$.

In general, the ``unbroken" vacua are shifted by the matter contribution, so
that the gauge symmetry at those vacua is actually broken. We can still refer
to these vacua as ``unbroken" in order to make explicit their origin in the
${\bf Z}_N$-vacua of the pure Yang--Mills theory. In fact, if the matter is
even slightly heavier than the compactification scale, $ML>1$, we can have  
 $\Delta \ll 1$ and the symmetry breaking effects (such as gauge boson masses)
are suppressed by a factor of $O(\Delta)$.

\subsec{Examples}

\noindent 

To illustrate these points, we consider some examples. First, matter fields
(fermionic or bosonic) in the fundamental representation of $SU(N)$.      
In this case, the weights are given directly by the $\nu^i$, so that the
potential is proportional to
$$
\sum_i V_M\left({2\pi \over N} \sum_k {\vec n}_k \right) =N V_M\left({2\pi \over N}
 \sum_k {\vec n}_k \right)  
\,.$$
For small values of the argument, we approximate the potential as in
\apot. Then, as  
 $\sum_k {\vec n}_k$ varies from $O(1)$ to $O(N)$, the lifted local minima
scan a band ranging from $O(1/N)$ to $O(N)$, times the mass quenching factor
$\Delta$,  with spacings  of order $\Delta/N$ at the bottom, and
only of order $\Delta$ at the top of the band.  

A second example is given by matter fields in the antisymmetric 
representation of $SU(N)$. 
In this case, weights are of the form $\nu^i + \nu^j$, with $i < j$. The
matter potential evaluated at the lattice of unbroken symmetry points is
proportional to
$$
\sum_{i<j} V_M\left({4\pi \over N} \sum_k {\vec n}_k \right) =\half N(N-1) 
\;V_M\left({4\pi \over N} \sum_k {\vec n}_k \right)   
\;.$$
 The same reasoning as before shows that
the lattice of $N^n$ points of unbroken $SU(N)$ symmetry is lifted to a
band with typical (bottom)
 spacing of $O(1) \,\Delta$ and ranging up to $O(N^2) 
\Delta$ energies. 
 If we consider the 
symmetric representation instead, we have to add the weights of the form
$\mu = 2\nu^i$,  introducing a finer structure of the  type already  
described for the fundamental representation. 

Our final example concerns the phenomenon of {\it planar equivalence}. It
refers to the situation when nonsupersymmetric theories still show vanishing
vacuum energy in the leading large $N$ approximation. In other words, when
the total number of perturbative degrees of freedom, counting polarizations
and various charges, including color, cancels among bosons and fermions. One
can generate one such example by  
 taking a supersymmetric theory, and replacing the
gauginos in the adjoint representation by Dirac fermions in the 
symmetric or antisymmetric representations, in such a way that the 
counting of degrees of freedom differs from the supersymmetric one only
at order $N$ (the so-called orientifold field theories \refs\rori, other 
examples can be found in \refs\rus). 
For instance, in an orientifold-type model, the potential is proportional
to   
\eqn\orip{
V_{\rm eff} ({\vec \phi}\,)_{\rm ori}
 \propto \sum_{i,j} V_0\left(\,(\nu^i - \nu^j)\cdot
{\vec \phi}\,\right) - 2 \sum_{i<j} V_0\left(\,(\nu^i + \nu^j ) \cdot
{\vec \phi}\,\right)
\;,}
up to terms of $O(N)$. 
In this case, the quenching effect of the matter mass is absent, so that
the local vacuum structure  of the full potential is significatively different
from that of the pure Yang--Mills model.    
The interesting fact about this example is the vanishing of the {\it averaged}
planar potential over the moduli space, i.e. the integral of      
$V_{\rm eff}$ over the toron valley is only of $O(N)$, despite the fact
that the generic scale of the potential is $O(N^2)$. To see this, we
write the flat connection in the basis of simple roots: ${\vec \phi} =
\sum_l {\vec c}_l \,\alpha_s^l$, with $\alpha_s^l = \nu^l - \nu^{l+1}\;,
l= 1, \dots, N-1$ the simple roots.  
The orientifold potential takes the form
\eqn\oriav{
V_{\rm eff} \left({\vec \phi}\,\right)_{\rm ori} =
 \sum_{i,j} \left[\,V_0\left(\shalf ({\vec c}_i - {\vec c}_{i-1} -
{\vec c}_j + {\vec c}_{j-1}
)\right) -  V_0\left(\shalf ({\vec c}_i -{\vec c}_{i-1} + {\vec c}_j -{\vec c}_{j-1} 
)\right)\right] + O(N)
\;.}
In this expression, we have neglected terms of $O(N)$ coming from  restrictions in  range of
the indices. When  averaging over the toron valley, the coordinates
${\vec c}_j$ become dummy integration variables, and fermionic terms cancel out
the bosonic ones by an appropriate  change of variables. 
Similar reasoning shows that the squared potential is at most of $O(N^3)$.
This means that potentials with the property of planar equivalence have no
  large ``plateaus" at heights of $O(N^2)$. 
Instead, plateaus stay at $O(N)$ and all $O(N^2)$ peaks and valleys are
relatively narrow.

\subsec{The general rules}

\noindent

In general, the condition for the unbroken  vacua to be lifted into a
broad band of energies is that the global $({\bf Z}_N)^n$ symmetry be
broken by the matter representations. At the same time, 
 the local stability properties
of the  pure Yang--Mills potential at those vacua   
should not be significantly upset, as is the case for matter potentials
generated by relatively large masses on the compactification scale. 

The misalignment responsible for
the lifting of the vacua finds its origin in the slight non orthogonality
of the basic weights $\nu^i$. If the expression of a given weight
$\mu$ in terms of the $\nu^i$ shows the same number of positive and negative
signs modulo $N$, then the terms proportional to $2\pi /N$ cancel out
in the argument of the scalar potential $V_M({\vec \xi})$.   The coefficient
of this term only depends on the representation, and not on the particular
weight within it, because all those differ by a integer linear combination
of roots, and the scalar product of roots with the $\nu^i$ leaves no
$O(1/N)$ residue.   

If $\nu^i$ are the weights of the defining fundamental representation (the
$N$), then 
$-\nu^i$ are weights of the conjugate representation (the ${\bar N}$).
 Any irreducible
representation can be found in the decomposition of the tensor product of
the $N$ and the ${\bar N}$ representations. Hence, the (mod $N$)
 number of $\nu^i$
minus the number of $-\nu^j$  in the expression for $\mu$ is  
the $N$-ality of the representation, i.e. the character under the
${\bf Z}_N$ center of the group.     
At the end, the distribution of cosmological constants at the
unbroken minima  ${\vec \phi}_0 = 4\pi
\sum_i {\vec n}_i \,\nu^i$ is given by (up to an additive cosmological
constant that is not determined by the potential)  
$$
 \Lambda ({\vec n}_i) = 
V_{\rm eff} \left({\vec \phi}_0\,\right) =\sum_R \CN_R \,(-1)^{F_R}\;
 {\rm dim} \,(R)\;
 V_{M_R} (2\pi {\vec K}_R /N)
\;,$$
 with  the integer vector ${\vec K}$,  
$$ 
{\vec K}_R = \eta_R \;\sum_k {\vec n}_k
\;,$$
and  $\eta_R$  the $N$-ality of the representation $R$.

Provided $ML$ is sufficiently large, each
 representation $R$ lifts the $N^n$ unbroken vacua into a band
of width 
$$
\Delta \Lambda = {\rm dim}(R)  \;\Delta
\;$$
 and level spacings in the range of  
$$
\delta\Lambda = {\rm dim} (R)  
 \;\left({\eta_R \over N}\right)^2 \,\Delta
$$
 at the bottom of the band.
 The previous results
for the fundamental representation are obtained by setting ${\rm dim}\,(R)=N$
and $\eta_R =1$, 
whereas those for the symmetric and antisymmetric representations use 
${\rm dim}\,(R) = \shalf N(N\pm 1), \;\eta_R =2$.  
In general, it is found that a discretuum of vacuum energies is favoured
by group-theory effects in the case of matter
 in the fundamental representation.
For higher-dimension matter representations, a quasi-continuous band 
 is still possible, but then it is entirely determined by the large
mass hierarchy $ML \gg 1$.

The $N$-ality is defined modulo $N$, and any representation
with vanishing $\eta_R$ fails to lift the local unbroken vacua. This does not
mean that these vacua remain unaltered, since local properties, such as
masses of particles, depend on the relative contribution of matter and gauge
terms (for example, fermion contributions tend to turn the minimum into a 
local maximum).

\newsec{Conclusions}

\noindent

In this note we have exhibited a simple example of landscape-like potential
generated dynamically  in purely field-theoretical terms.
 It has few continuous adjustable
parameters, owing its peculiar features to group-theory effects. 
  The effective potential
for theories whose matter sector
  breaks the global group of central conjugations  
shows a very intricate pattern of peaks and valleys. In particular, local
vacua of unbroken $SU(N)$ symmetry get lifted across a band  
with potentially small level spacing. A 
degeneracy of $O(N^{n-1})$ vacua remains, as a consequence of the
hypercubic symmetry of ${\bf T}^n$. The vacuum splittings depend on  
the contribution of matter representations, and  can be controlled by
the quenching factor $\exp(-ML)$ induced by large masses. Hence, a large
mass hierarchy between the matter representations and the compactification
scale  produces a finer band of cosmological constants.     

The main limitation of these considerations is their perturbative nature.
In interesting $d=4$ situations, the higher-dimensional Yang--Mills theory
is necessarily ill-defined in the ultraviolet. A UV completion is needed
at some threshold scale $\Lambda_{\rm UV} \gg 1/L$. The accuracy of our description
of the effective potentials depends on the ability to maintain a large
hierarchy between this UV scale and the compactification scale. The
dimensionless  't Hooft
coupling  normalized at the compactification scale $\lambda = g^2 (L) N \,
L^{4-d}$ controls our approximations, valid for $\lambda \ll 1$, with
corrections suppressed by fractional powers of $\lambda$.    

Given the relationship between higher-dimensional theories and quiver
models, it would be interesting to check if 
analogous potentials with scanning properties can be derived in the
framework of ``deconstruction" \refs\rdec. 
Possible  applications of these results to Kaluza--Klein model building   
 will not be considered here. We simply note that any
implementation of this construction in string models with higher-dimensional
wrapped branes (the natural arena to reproduce these features) requires
that the gauge spectrum on the world-volume  remains  non-supersymmetric   
much above the compactification scale. This means that such models must
have high-scale supersymmetry breaking, at least in this sector.
The additional requirement that $\lambda \ll 1$ implies that such potentials
are likely to be useful only for the modeling of  ``hidden sectors" (for
other recent ``landscape" constructions in the open string sector, see 
\refs\rmat).

 At any rate,  
 for the ``discretum" of vacuum energies to have any phenomenological
interest, one would 
 need to drastically reduce the energy gap between local vacua.
The group-theoretical gaps are smaller for matter in the fundamental 
representation. Further suppression 
 might be achieved by constraining the matter representations to be
very massive, $ML \gg 1$. The global overall scale of the potential (both
for matter and gauge contributions) can be reduced by 
 placing the branes at the bottom of a long throat in 
 the context of warped compactifications.

\vskip 1cm

{\bf Acknowledgements}

\noindent

We thank Karl Landsteiner for useful comments. 
The work of C.H. was partially supported by FPU grant AP2002-0433 from MEC-Spain.
 The work of J.L.F.B. was partially supported by MCyT
 and FEDER under grant
BFM2002-03881 and
 the European RTN network
 MRTN-CT-2004-005104.

\listrefs

\bye